\documentclass{aa}
\usepackage{graphicx}
\begin{document}
\title{Short-term spectroscopic monitoring of two cool dwarfs with strong
       magnetic fields
\thanks{Partly based on observations obtained at the European Southern Observatory
at La Silla, Chile in programs 078.C-0161(A) and 078.C-0161(B), and partly based on 
observations collected at the Centro Astron\'omico Hispano Alem\'an /CAHA) at Calar Alto,
operated jointly by the Max-Planck-Institut f\"ur Astronomie and the the Insituto de
Astrof\'\i sica de Andaluc\'\i a (CSIC).}}
\authorrunning{Guenther et al.}
\titlerunning{Spectroscopic monitoring of cool dwarfs}


   \author{E.W. Guenther\inst{1}
          \and
           M.R. Zapatero Osorio\inst{2}
          \and
           A.Mehner\inst{3}
          \and
           E.L. Mart\'\i n\inst{2}
             }

\offprints{Eike Guenther, \email{guenther@tls-tautenburg.de}}

     \institute{Th\"uringer Landessternwarte Tautenburg,
                Sternwarte 5, D07778 -- Tautenburg, Germany
               \and
                Instituto de Astrof\'\i sica de Canarias,
                C/V\'\i a L\'actea, s/n,   
                E38205 -- La Laguna (Tenerife), Spain
               \and
                Department of Astronomy, University of Minnesota,
                116 Church St. SE,  Minneapolis, MN 55455, USA
              }

   \date{Received 19.5.2008; accepted 11.12.2008}

\abstract{There is now growing evidence that some brown dwarfs (BDs)
  have very strong magnetic fields, and yet their surface temperatures
  are so low that the coupling is expected to be small between the
  matter and the magnetic field in the atmosphere. In the deeper layers, 
  however, the coupling is expected to be much stronger.}
{This raises the question of whether the magnetic field still leads to
  the formation of structures in the photosphere and of a solar-like
  chromosphere and corona.}
{We carried out a spectroscopic monitoring campaign in which we
  observed ultracool dwarfs that have strong magnetic fields: the BD
  LP944-20 and 2MASSW\,J0036159+182110. The objects were monitored
  over several rotation periods spectroscopically. LP944-20 was
  observed simultaneously in the optical and in the near infrared
  regime, 2MASSW\,J0036159+182110 only in the infrared.  From the
  spectra, we determined the temperature of the objects in each spectrum,
  and measured the equivalent width in a number of diagnostically
  important lines. Temperature variations would indicate the presence
  of warm and cold regions, variations in the equivalent widths of
  photospheric lines are sensitive to the structure of cloud
  layers, and $H_{\alpha}$ is a diagnostic for chromospheric
  structures.}
{Both dwarfs turned out to be remarkably constant.  In the case of
  LP944-20, the $T_{eff}$-variations are $\leq 50$\,K, and the
  rms-variations in the equivalent widths of $H_{\alpha}$ small. We
  also find that the equivalent widths of photospheric lines are
  remarkably constant. We did not find any significant variations in
  the case of 2MASSW\,J0036159+182110 either.  Thus the most important
  result is that no significant variability was found at the time of
  our observations. We find that H$\alpha$-line is in emission
    but the equivalent width is only $-4.4\pm0.3$ \AA . 
    When comparing our spectra with spectra taken over the past 
    11 years, we recognize significant changes during this time.}
{We interpret these results as evidence that the photosphere of these
  objects are remarkably homogeneous, with only little structure in
  them, and despite the strong magnetic fields.  Thus, unlike
  active stars, there are no prominent spots on these objects.}


\maketitle


\section{Introduction}

Brown dwarfs (BDs) are objects that are not massive enough to sustain
stable thermonuclear fusion of hydrogen at their centers but are
distinguished from gas-giant planets by their ability to burn
deuterium. Among many other things, these objects are interesting
because their properties place them some where between planets
and stars. In M-stars, strong chromospheric emission lines are
originating from an active chromosphere, and thus provide evidence of
correspondingly strong magnetic fields generated by a stellar
dynamo. Basri \& Marcy (\cite{basri95}) studied the relation between
$v\sin\,i$ and the strength of $H_{\alpha}$ for very low-mass stars
and one BD-candidate. Surprisingly, they find that the most rapid
rotator of their sample exhibits no emission in $H\alpha$.  In a
subsequent study, Mohanty \& Basri (\cite{mohanty03}) find a drastic
drop in activity and a sharp brake in the rotation-activity
connection. The $H\alpha$ emission levels in very late type dwarfs are
much lower than in earlier types and often undetectable, in spite of
very rapid rotation.  The photometric variability of L-type BDs has
henceforth been interpreted in terms of clouds and weather, as on
planets (Morales-Calder{\'o}n et al. \cite{morales} and the reference
therein).  At first glance, magnetic fields and spots seemed to
be unimportant for the structure of the atmospheres of old BDs.

The detection of X-ray emission and large flares indicating the
presence of a corona and of strong magnetic fields in at least some
BDs changed the picture dramatically (Liebert et al. \cite{liebert03};
Burgasser \& Putman \cite{burgasser05}; Preibisch \& Zinnecker
\cite{preibisch02}; Preibisch et al. \cite{preibisch05}; Ozawa et al.
\cite{ozawa05}; Fleming et al. \cite{fleming03}). The question thus
arises whether BDs are like M-dwarfs in this respect. For both active
and inactive stars, there is a correlation between the X-ray and the
radio emission of the corona, which works for over 10 orders of
magnitude in activity level.  It was thus very surprising when Berger
et al.  (\cite{berger05}) discovered that \object{LP944-20} is 4 to 5
orders of magnitude too bright in the radio regime. The same phenomena
has also been observed for a few other BDs. The coronae of these
objects thus must be quite different from those of normal stars!
Observation at 8.46 GHz with the VLA of the old BD
\object{2MASSW\,J0036159+182110} (from now on called
\object{2M0036+1821}) imply a magnetic field strength of $\sim$ 175 G
at about two radii above the surface of the object (Berger
\cite{berger06}).  The field strength at the surface must be $\geq 1
kG$.  Thus, it is now clear that at least these old BDs have strong
magnetic fields indeed, but are these BDs just like active stars?

It is possible that not only do the coronae differ from those of stars
but also the topology of the magnetic field itself. As shown by Dobler
et al.  (\cite{dobler06}), fields of fully convective objects (like
BDs) are expected not to be concentrated in small spots but to be
distributed on a global scale.  Chabrier \& K\"uker
(\cite{chabrier06}) find that the field for fully convective objects
should be generated by an $\alpha^2$ dynamo: the fields on a large
scale, and are non-axis symmetric. In this respect it is interesting
to note that Zeeman-Doppler imaging observation of a fully convective,
rapidly rotating 0.28 $M_\odot$-star shows a strong, large-scale, but
axisymetric field (Donati et al. \cite{donati06}). This shows that
more observations and theoretical work are needed to
understand the fields of fully convective objects.  That
brown dwarfs are rapid rotators (Zapatero Osorio et
al. \cite{zapatero06}) must, however, be related to the absence of
any winds that are similar to the solar-wind that could brake these objects.
Possibly the absence of such winds is related to the topology of the
magnetic field, rather than to its strength. A solar magnetic field
topology is only expected for very old, very massive BDs, which have
conductive cores. The other difference for the solar-like stars is the
low temperature of the atmosphere, resulting in a low degree of
ionization in the atmosphere, which in turn lead to a very low
degree of coupling between the magnetic field and the atmosphere. The
coupling between the gas and the magnetic field is usually described
in terms of the Reynolds number $R_m=l\upsilon/\eta$ (where $l$ is a
length scale, $\upsilon$ a velocity scale, and $\eta$ the magnetic
diffusivity (Priest \cite{priest82}).

Following a suggestion by Meyer \& Meyer-Hofmeister (\cite{meyer99}),
Mohanty et al. (\cite{mohanty02}) studied the conductivity of the
atmospheres of late M and L dwarfs. They find that the atmospheres
have very high electrical resistivities because they are predominantly
neutral. For example, ionization fraction at $\tau_{J-band}=1$ is only
between $10^{-5.5}$ and $10^{-7}$ for late M down to L-dwarfs.  On the
other hand, underneath the surface the temperature increases rapidly,
and the conductivity of the matter is expected to be high. Thus in the
interior of the object, the magnetic field will interact with the
convection. If this is the case, we might speculate that such effects
might then lead to the formation of hotter and cooler regions at the
surface, if the convective energy transport from the interior is
affected by the magnetic field. In summary, at least some BDs have
strong magnetic fields. We also know that they have flares, a
chromosphere, and a corona, but what the effects of the fields on the
atmosphere are is not known.

   \section{The two BDs}

An ideal object for studying the effects of magnetic fields in BDs is
\object{LP944-20}.  Using measurements of the equivalent widths of the
LiI\,6708-line, together with models, Tinney (\cite{tinney98})
estimate, the age of \object{LP944-20} to be between 475 and 650
Myrs., and he estimates the mass to be between 0.056 and 0.064
$M_\odot$.  Pavlenko et al. (\cite{pavlenko07}) finds a
two-orders-of-magnitude higher abundance of lithium compared to the
older determination, corresponding to the expected primordial
abundance, which would imply that the object is very young. However,
this somehow contradicts the result by Johnas et al.
(\cite{johnas07}) that the lithium of BDs with similar mass is already
reduced at an age of only 1 Myr.  Also, because the object does not
have a disk (Apai \cite{apai02}), it is unlikely that it is very young.
Ribas (\cite{ribas03}) finds that \object{LP944-20} is a member of the
Castor moving group. They both conclude that the age is $320\pm80$ Myr, and
the mass between 0.049 and 0.055 $M_\odot$.  The parallax is
$201.4\pm4.2$ mas, which corresponds to a distance of $5.0\pm0.1$ pc
(Tinney \cite{tinney96}).

\object{LP944-20} is not only relatively young and nearby, it is also
very active. Large flares have been observed in the X-ray regime (e.g.
Rutledge et al. \cite{rutledge00}). Berger et al.  (\cite{berger05})
found that it is a non-thermal radio source at GHz-frequencies,
indicating the presence of a magnetic field. Interestingly, while the
object is very bright at GHz-frequencies, the quiescent X-ray flux has
not been detected yet. The upper limit is $log(L_x/L_{bol}) \leq
-6.28$ (Mart\'\i n \& Bouy \cite{martin02}). Using the relation
between X-ray luminosity and the flux at GHz-frequencies of stars,
\object{LP944-20} is in fact more than 5 orders of magnitude too
bright in the radio regime (Berger et al. \cite{berger06}). From the
radio observations, the authors estimate a magnetic field strength of
135 G during a flare, and $<$ 95 G in quiescence at a distance of
about one BD-radius from the surface. The rotation velocity 
$v\,sin\,i$ is $28\pm4$ $km\,s^{-1}$, which implies a rotation period of
$\leq 4.5$ hours (Guenther \& Wuchterl \cite{guenther03}).

If \object{LP944-20} were an active star, we would expect it to have
large spots. If there were spots on the surface of the object, we
would see temperature variations. Interestingly, Tinney \& Tolley
(\cite{tinney99}) derive temperature variations of only
$\sim20\,K$. However, they monitored the object only for 1.5 hours,
and as will be discussed below, the TiO-lines are not ideal for
determining the temperature for an object with a spectral class of
M9V.  In a previous campaign, we took 15 spectra with UVES of
\object{LP944-20} and found radial velocity variations in the optical
regime but none in the infrared regime (Mart\'\i n et
al. \cite{martin06}).

Another suitable object for this project is \object{2M0036+1821}. It
is the brightest, very cool dwarf at GHz-frequencies in the northern
hemisphere, and it also has a strong magnetic field (Berger
\cite{berger06}).  Similar to \object{LP944-20}, \object{2M0036+1821}
dramatically violates the stellar relation for the flux ratio between
the radio and X-ray regime as it is a factor $10^4$ to $10^5$-times
too bright in the radio regime. Unfortunately, the age, hence the
mass, of \object{2M0036+1821} is not known. Given a spectral type of
L3.5 (Kirkpatrick et al.\cite{kirkpatrick00}), $L4\pm1$ (Cushing et
al. \cite{cushing06}), or $L5\pm1$ (Reid et al. \cite{reid00}), and
the fact that we do not know the age of the object we cannot be
certain whether it is a brown dwarf or a very low-mass star.  We thus
call it an ultra cool dwarf.  The parallax is $114.5\pm8$ mas, which
implies a distance of $8.7\pm0.1$ pc (Dahn et al. \cite{dahn02}).

The $v\,sin\,i$ measurement $36.0\pm2.7$ $km\,s^{-1}$ derived by
Zapatero Osorio et al. (\cite{zapatero06}) and the quasi-period found
by Berger et al. \cite{berger05} indicate that the rotation period is
possibly about 3 hours.  Maiti (\cite{maiti07}) detected significant
photometric variations in the R and I-bands. The $\sigma_{rms}$ is 0.01
and 0.03 mag in the two bands, respectively. Curiously, when the
amplitude of the variations is larger in the R-band, the amplitude in
the I-band appears to be smaller, and when it is larger in the I-band,
it appears to be smaller in the R-band. M{\'e}nard et
al. (\cite{menard02}) and Zapatero Osorio et al. (\cite{zapatero05})
reported on the detection of optical (R and I) linear polarization in
\object{2M0036+1821}, which the authors ascribed to the presence of
atmospheric clouds.  Sengupta \& Kwok (\cite{sengupta05}) reproduce
the polarization observations of \object{2M0036+1821} at three
different wavelengths by using a rotational velocity of 15
$km\,s^{-1}$ and a grain size of 0.46 micron. According to the
most recent $v\,sin\,i$ measurements, the rotation speed of
\object{2M0036+1821} is a factor of 2 higher. If rotational velocity
is increased, grain size should become smaller for the same
polarization degree.

   \section{Strategy of the observations}

The aim of this work is to find out whether strong magnetic fields
lead to surface features or not.  To answer this question, we carried
out a time series of spectroscopic observations of \object{LP944-20}
and 2M0036+1821.  From an ana\-lysis of optical and X-ray data of
several BDs, Tsuboi et al. (\cite{tsuboi03}) found an indication of a
relation between the X-ray brightness and the strength of the
$H_{\alpha}$-line, which they interpret as a signature that the
chromosphere (traced by $H_{\alpha}$) is heated by the corona. By
monitoring the strength of the $H_{\alpha}$-line, we thus might trace
the structures not only in the chromosphere but in the corona as
well. The second important parameter is the surface temperature, which
can be determined by deriving the PC3-coefficient (Mart{\'{\i}}n et
al. (\cite{martin99}); ratio of the fluxes in the 823.5 to 826.5 nm,
versus the fluxes in the 754.0 to 758.0 nm- band.).

Apart from monitoring the $H_{\alpha}$-line and possible changes in
the temperature, the spectra also allow us to monitor any possible
changes in the cloud structure. Allard et al. (\cite{allard01})
calculated spectra for a dense grid of $T_{eff}$-values.  These
calculations show that clouds greatly reduce the equivalent widths of
photospheric lines and that the effects of clouds are more pronounced
in the J-band than in the K-band.  New calculations by Burrows et al.
(\cite{burrows06}) show that not only the presence of clouds can be
inferred from IR spectra but also the distribution of condensables in
the atmosphere.

   \section{LP944-20}

   \subsection{EFOSC2 observations of LP944-20}

\begin{figure}[h]
\includegraphics[width=0.35\textwidth, angle=270]{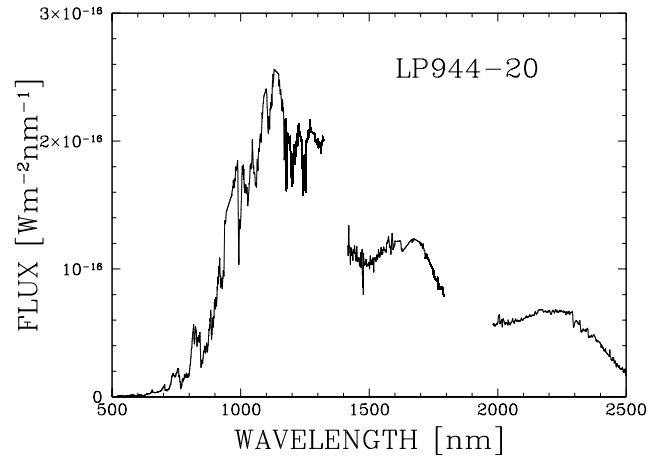}
\caption{Average spectrum of \object{LP944-20}.}
\label{fig01}
\end{figure}

A time series of spectra of \object{LP944-20} was taken with the
EFOSC2 (ESO Faint Object Spectrograph and Camera Version 2), which is
operated at the ESO 3.6-m telescope at La Silla, Chile.  Grism no. 5
was used for the observations.
This grism covers the wavelength
region between 520 nm and 935 nm. The resolving power is $\lambda
/\Delta \lambda$=450 with the 1.2 arcsec slit. The slit was put
at the parallactic angle to minimize light-loss. Standard IRAF
tasks were used for bias subtraction, flat-fielding, sky-subtraction,
extraction, and wavelength calibration of the spectra. Unfortunately,
EFOSC2 has very strong problems with fringes in the red part of the
spectrum.  The fringes depend on the telescope position, so that
flat fields had to be taken during the nights. Although the amplitude
of the fringes were greatly reduced after flat-fielding, the remaining
fringes were still so strong that the wavelength region between 720
and 935 nm was affected. The spectra were flux-calibrated using the
standard star \object{EG21}. \object{LP944-20} was observed in the
night from 29 to 30 November 2006, in which 14 spectra were taken
and in the night from 30 November to the 1 of December 2006 in
which 16 spectra were taken. An average spectrum is shown in
Fig.\,\ref{fig01} which also includes all the SOFI spectra. A log
of the observations of \object{LP944-20} is shown in
Table\,\ref{tab1}.

\begin{table*}
\caption{Observing log}
\begin{tabular}{ccccc}
\hline \hline
object & wavelength-  & HJD           & date & UT  \\
       & region       & start - end   &      & start - end \\
       & [nm]         & 2450000+      &      &  \\
\hline
2M0036+1821 & 1070-1340 & 3987.3660-3987.4666 & 8. Sep 2006  & 20:40 - 23:05 \\
            & 1580-2400 & 3987.3770-3987.4438 & 8. Sep 2006  & 20:53 - 22:32 \\
2M0036+1821 & 1070-1340 & 3987.6123-3987.6924 & 9. Sep 2006  &  2:35 -  4:30 \\
            & 1580-2400 & 3987.6208-3987.7009 & 9. Sep 2006  &  2:47 -  4:42 \\
2M0036+1821 & 1070-1340 & 3988.5699-3988.5891 & 10. Sept 2006 &  1:33 - 2:01 \\
            & 1580-2400 & 3988.5785-3988.5804 & 10. Sept 2006 &  1:46 - 1:49 \\
2M0036+1821 & 1070-1340 & 3988.6554-3988.7101 & 10. Sept 2006 &  3:36 - 4:55 \\
            & 1580-2400 & 3988.6654-3988.7018 & 10. Sept 2006 &  3:51 - 4:43 \\
2M0036+1821 & 1070-1340 & 3989.4429-3989.6953 & 10./11. Sept 2006 & 22:31 -  4:34 \\
            & 1580-2400 & 3989.4519-3989.6858 & 10./11. Sept 2006 & 22.43 -  4:20 \\
LP944-20    &  520-935  & 4069.5794-4069.8551 & 30. Nov 2006 & 1:50 - 8:27  \\
            &  938-1646 & 4069.5475-4069.8607 & 30. Nov 2006 & 1:04 - 8:35  \\
            & 1502-2538 & 4069.5558-4069.8697 & 30. Nov 2006 & 1:16 - 8:48  \\
LP944-20    &  520-935  & 4070.5919-4070.8578 & 1. Dec 2006  & 2:08 - 8:31  \\
            &  938-1646 & 4070.5530-4070.8259 & 1. Dec 2006  & 1:12 - 7:45  \\
            & 1502-2538 & 4070.5641-4070.8349 & 1. Dec 2006  & 1:28 - 7:58  \\
\hline\hline
\end{tabular}
\label{tab1}
\end{table*}

\begin{figure}[h]
\includegraphics[width=0.35\textwidth, angle=270]{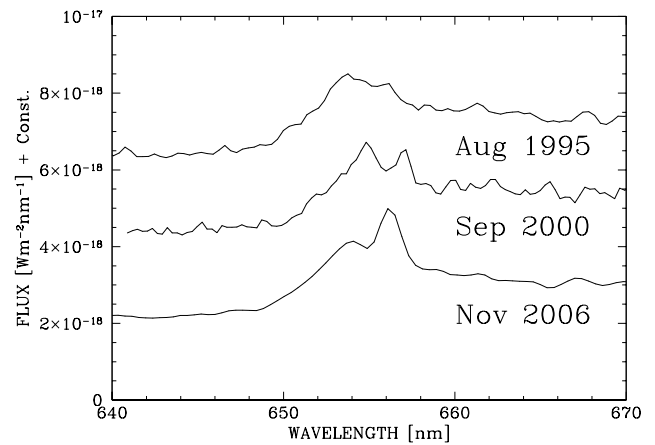}
\caption{Comparison of the spectrum of \object{LP944-20} taken by us
  on 2006 Nov. 30 and Dec. 1 (EFOSC), with the spectrum taken by
  E. Mart{\'{\i}}n 2000 Sep. 28 (taken with the WHT) and with the one
  published by C. Gelino, D. Kirkpatrick, A. Burgasser (\cite{gelino})
  taken 1995 August 12 (taken with the Blanco telescope). The spectrum
  taken in 2006 shows a much stronger emission-line component of
  $H_{\alpha}$ than the older spectra.}
\label{fig02}
\end{figure}

\begin{figure}[h]
\includegraphics[width=0.35\textwidth, angle=270]{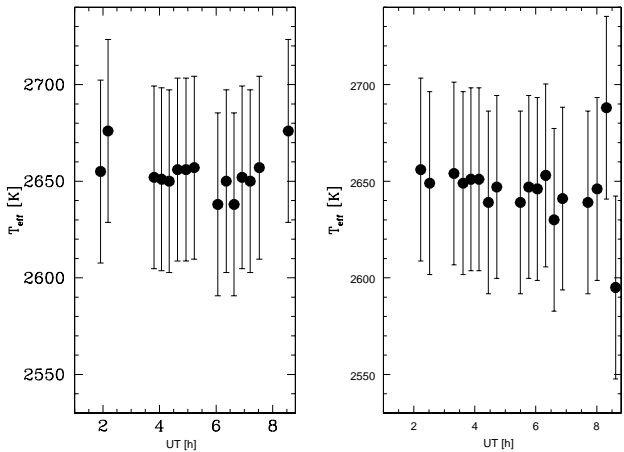}
\caption{Temperature measurements of \object{LP944-20}.  The left
  panel is for the first observing night, the right one for the
  second. The errors were derived from the accuracy of the
  flux-calibration. Temperature variations are $\leq 30\,K$.}
\label{fig03}
\end{figure}

\begin{figure}[h]
\includegraphics[width=0.35\textwidth, angle=270]{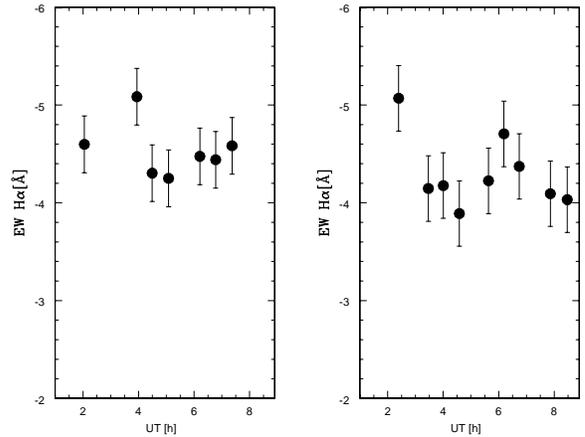}
\caption{Measurements of the equivalent width of $H_{\alpha}$ of
  \object{LP944-20}. The left panel is for the data taken on 2006 Sept
  30, the right one for the data taken of 2006 Dec 1. The
  amplitudes of the variations are relatively small.}
\label{fig04}
\end{figure}

   \subsection{Results from the optical spectroscopy of LP944-20}

As a first step, we determined the temperature of \object{LP944-20}
using the PC3-coefficient from Mart{\'{\i}}n et al. (\cite{martin99})
and from this the spectral types.  This coefficient allows the
spectral types to be determined for objects in from M3 to L5.  For
converting the spectral type into $T_{eff}$, we used the values given
in Burgasser \& Kirkpatrick (\cite{burgasser06}), Burgasser
(\cite{burgasser07}), and Nakajima (\cite{nakajima04}).  Determining
the absolute temperature of an object with a late-spectral type is
difficult, because it involves not only the measurement of the ratio
of the fluxes in two spectral regions but also the uncertainty of the
model spectra. It is thus not surprising that the errors can be as
large as several hundred K. However, our main focus is to detect
temperature variations. Detecting these is much easier. For example,
it is known that the PC3-coefficient changes linearly with temperature
between about spectra type M3V and M8V (about 2400 to 3300 K). Within
this region, we can thus convert even a small variation in the
PC3-coefficient into a variation of the temperature. Since even the
smallest variation of the temperature will lead to a change in the
PC3-coefficient, we are also able to detect the small variations of
the temperature. The accuracy with which the variation of the
PC3-coefficient can be detected is given by the accuracy of the
flux-calibration in the 754.0 to 826.5 nm regime, which is dominated
by the variable extinction in the Earths atmosphere. Of course we
always used the observations of the flux-standard \object{EG12} which
were closest to the observations \object{LP944-20}. For estimating the
accuracy of the flux-calibration, we simply used the other
observations of \object{EG12} and derived the errors from the
difference. We thus overestimated the errors somewhat. We find an
upper limit of the error of $\leq 5.4\%$ in this wavelength
regime. This corresponds to an error of $\pm 47\,K$ for
\object{LP944-20}. The results are shown in Fig.\,\ref{fig03}.  While
we are mainly interested in temperature variations, it is interesting
to note that $T_{eff}$ values derived match the temperature obtained
by Mohanty \& Basri (\cite{mohanty03}) using HIRES spectra very well.

The left panel of Fig.\,\ref{fig03} is for the observations taken
2006 Sept 30, the right one for the data taken of 2006 Dec 1.
There are no significant variations in the temperature. Since $\sigma$
is only 15 K, it seems that we slightly overestimated the errors. From
Fig.\,\ref{fig03}, we estimate that the temperature variations are
$\leq 30\,K$.

Tinney \& Tolley (\cite{tinney99}) used the TiO-bandhead for
determining the temperature variations. While this coefficient works
well up to a spectral type M6V, it is also affected by the
distribution of condensables in the atmosphere at later spectral
types. Nevertheless, to be able to compare our results with those of
Tinney \& Tolley (\cite{tinney99}), we determined the ratio of the
fluxes in the 704.2 to 704.6 nm -band and in the 712.6 to 713.5 nm
-band (e.g. $TiO\,5$-coefficient Reid et al. \cite{reid95}). If we
express the $TiO\,5$-coefficient in temperature, we find an upper
limit of $\leq 15\,K$. Thus, our results agree with the previous ones.

\begin{figure}[h]
\includegraphics[width=0.35\textwidth, angle=270]{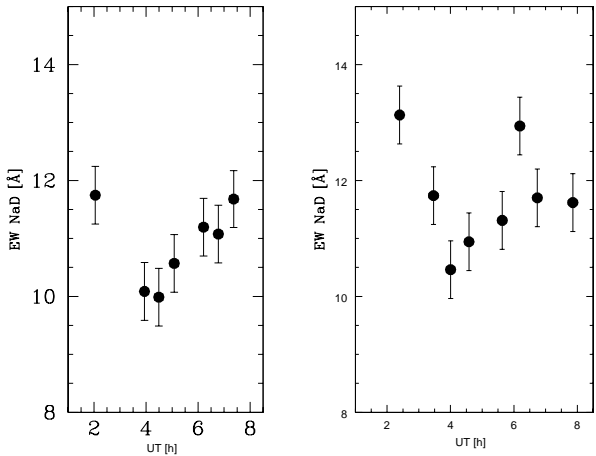}
\caption{Measurements of the equivalent width of NaD of
  \object{LP944-20}. The two panels correspond to the two observing
  nights.}
\label{fig05}
\end{figure}

\begin{figure}[h]
\includegraphics[width=0.35\textwidth, angle=270]{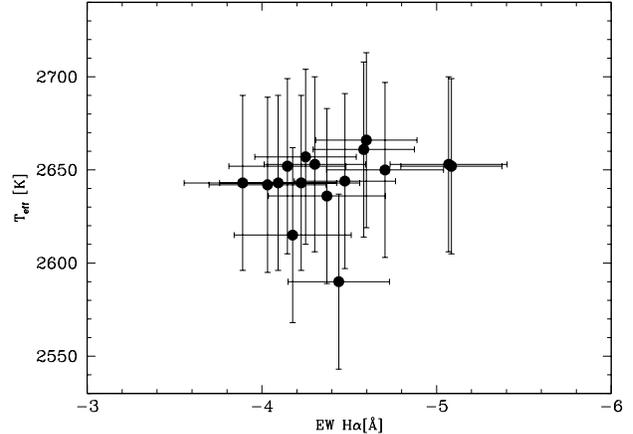}
\caption{Measurements of the equivalent width of $H_{\alpha}$ plotted
  against the temperature for \object{LP944-20}. As expected,
  there is no correlation.}
\label{fig06}
\end{figure}

Fig.\,\ref{fig02} shows the spectrum $H_{\alpha}$-region taken by us
(Nov 2006), together with a spectrum taken with the red arm of the
ISIS spectrograph on the WHT by Mart{\'{\i}}n in September 2000, and a
spectrum taken in August 1995 with the CTIO 4m Blanco telescope
(Gelino, Kirkpatrick \& Burgasser \cite{gelino}).  In these 11 years,
the $H\alpha$-emission of \object{LP944-20} has noticeably changed. In
our spectra $H_{\alpha}$ is clearly in emission.  Is this component
variable on shorter time-scales? The average of the
(pseudo-)equivalent width (pEW) of $H_{\alpha}$ is $-4.5\pm0.3\,\AA$
in the first night and $-4.3\pm0.3\,\AA$ in the second night.  Looking
at Fig.\,\ref{fig04} one might speculate that there could be
flare-like event in the second night, however, spectra of higher
resolution would have been necessary in order to find out, whether
this was a flare or not. In either case, we did not detect any strong
flare-activity in H$\alpha$.  In Fig.\,\ref{fig06} the $T_{eff}$ is
plotted against the equivalent width of $H_{\alpha}$. As expected
there is no correlation, and $T_{eff}$ and $H_{\alpha}$ can best be
described as almost constant.  For completeness, we also show in
Fig.\,\ref{fig05} the measurements the NaD line.

   \subsection{SOFI observations of LP944-20}

Simultaneous with the optical spectroscopy, we also obtained infrared
spectroscopy with SOFI, which is mounted on the NTT. During the nights
we alternated between the BLUE and the RED low-dispersion grism.  The
BLUE grisms gives a resolution of $\lambda /\Delta \lambda$=930 and
the RED grism one of 980 with the 0.6 arcsec slit used for the
observations. The BLUE grism covers the wavelength region from 938 to
1646 nm, and the RED grism 1502 to 2538 nm. Thus, with both grisms, we
covered the J, H, and K-bands. Standard IRAF tasks were used for bias
subtraction, flat-fielding, sky-subtraction, extraction, and
wavelength calibration of the spectra. A log of the observations of
\object{LP944-20} is shown in Table\,\ref{tab1}. The spectra were
flux-calibrated using B4V star \object{HD955}, which was observed each
night.  Along with \object{LP944-20}, we took spectra of
\object{DENIS-P\,104814.9-395604} and \object{LP647-13}, both of which
are supposed to have the same spectral type as \object{LP944-20}.

   \subsection{Results from the infrared observations of LP944-20}

\begin{figure}[h]
\includegraphics[width=0.35\textwidth, angle=270]{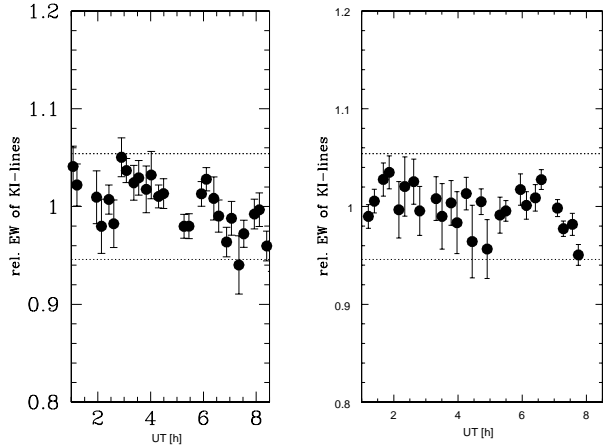}

\caption{Studies of the changes of the EW of the potassium
    lines at 1168, 1177, 1243, and 1254 nm of \object{LP944-20}. Since
    the four lines have about the same EW we averaged them.  Rather
    than showing the average EWs, we normalized the values.  The
  dashed lines are the averaged 3$\sigma$-errors of the
  measurements. 
}
\label{fig07}
\end{figure}

As mentioned above, the presence of clouds in the atmosphere of the BD
reduces the equivalent widths of the absorption lines, especially in
the J-band. Changing cloud patterns and changes in the position of the
cloud layers also lead to changes in the equivalent widths of spectral
lines. Since we already know that the temperature of \object{LP944-20}
is almost constant, by measuring the equivalent widths of photospheric
lines we thus plumb the cloud layers of it.  We choose in the J-band
the following lines: The Na\,I doublet (1138 and 1141 nm), another
Na\,I line (1268 nm), the two potassium doublets (1168, 1177 nm, and
1243, 1254 nm), and two FeI lines (1189 and 1197 nm). In the H-band
there are basically only the K\,I doublet line at 1517 nm, and the
FeH-band heads at 1583, 1591, and 1625 nm. In the K-band there are the
Ca\,I triplet at 1980 nm and the NaI doublet at 2206 and 2209
nm. Additionally, there are the CO band heads. Because the four
potassium lines are the strongest lines in the J-band, these lines
give the highest accuracy.  After determining the equivalent widths of
the potassium lines in each spectrum, we divided the values obtained
by the average equivalent width of each line. In this way we obtained
the normalized variations in the equivalent width for each potassium
line. Fig.\,\ref{fig07} shows the variation of the average EW of the
four lines at 1168, 1177, 1243, and 1254 nm. Rather than show the
average EW of the four lines, we normalized the data to one. A value
of 0.9, for example, means that that on average the EW of the four
lines are 10\% smaller than usual.  None of the measurements deviates
by more than 3$\sigma$ from unity.  Thus, there are no significant
variations of the equivalent width. The rms-variations of the
K\,I-lines of \object{LP944-20} are $\leq 2.0 \%$, and the variance of
the measurements is 2.8\%.  From the absence of significant variations
in the equivalent width, we conclude that the atmosphere of
\object{LP944-20} must be very homogeneous indeed.

   \section{2M0036+1821}

   \subsection{MAGIC observations of 2M0036+1821}

Using MAGIC and the Resin-Replica-Grism it is possible to take spectra
of the H and K-band simultaneously. Another setting is then required
for the J-band. The J-band grism covers the region from 1070 to 1340
nm, and the H\&K band grism the 1580 to 2400 nm region. The resolution
is $\lambda/\Delta \lambda$=450 in the J band, and 350 in the H and K
bands. We monitored \object{2M0036+1821} for three nights. Each night,
we observed the standard star \object{88 Peg} three or four times.
Unfortunately, the observing run was plagued by clouds so that in the
first two nights observation were only possible for about two hours,
and in the last for about 8 hours.  A log of the observations of
\object{2M0036+1821} is shown in Table\,\ref{tab1}. During all three
observing nights, clouds interfered with the observations, making the
analysis of the data complicated.  Standard IRAF tasks were used for
bias subtraction, flat-fielding, sky-subtraction, extraction, and
wavelength calibration of the spectra.

   \subsection{Results from the infrared observations of 2M0036-1821}

\begin{figure}[h]
\includegraphics[width=0.35\textwidth, angle=270]{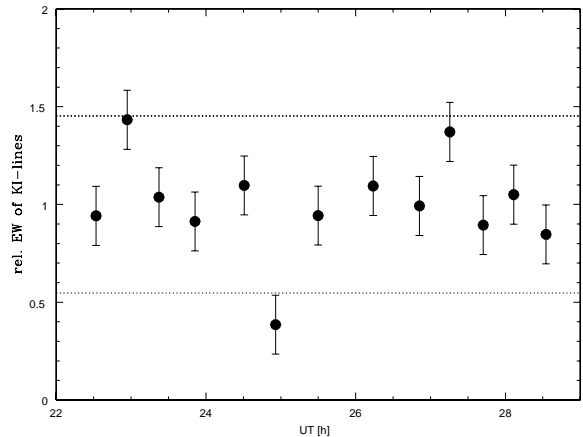}
\caption{Similar to Fig.\,\ref{fig07} but for the KI lines at 1243
  and 1254 nm and for \object{2M0036+1821} in the night
  10.-11. Sept. 2006 (JD 2453926.5). The two dotted lines are 
  3$\sigma$ above and below the average.}
\label{fig08}
\end{figure}

\begin{figure}[h]
\includegraphics[width=0.35\textwidth, angle=270]{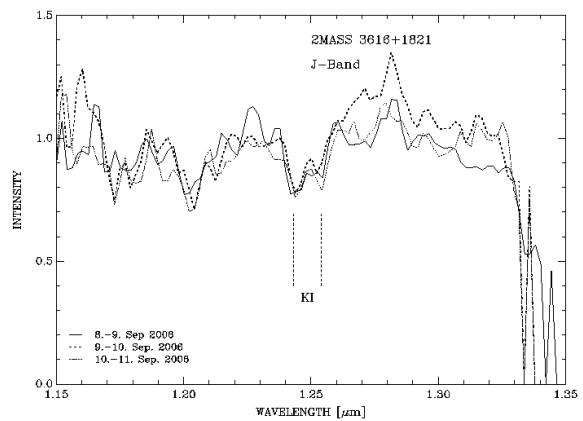}
\caption{J-band spectrum of \object{2M0036} for the
  three different nights.}
\label{fig09}
\end{figure}

Fig.\,\ref{fig08} shows the measurements of the relative change in the
equivalent width of the two potassium (KI) lines at 1243 nm and 1254
nm.  Again, none of the measurements deviate by more than 3$\sigma$
from unity. There are no significant variations in the equivalent
widths of these lines. The errors of the measurements for
\object{2M0036+1821} are, however, much larger than those for
\object{LP944-20}, as the variance is only $\leq 25 \%$.
Fig.\,\ref{fig09} shows the average J-band spectra taken those three
nights. Again, we do not see significant variations of the KI lines at
1243 and 1254 nm in these three nights.  Using the K-band spectra and
the $H_2O-A$, and $H_2O-C$ coefficients from Burgasser et
al. (\cite{burgasser02}) we determined the ``spectral type'' of each
spectrum, and then converted the spectral type into a
temperature. Because of the less than perfect observing conditions,
these measurements are not very accurate, and we find an upper limit
for any possibly variations of $\pm 200$ K.

   \section{Discussion and conclusions}

Because the temperature of the photosphere of stars is high, the
magnetic field interacts with the plasma forming the well-known spots
of active stars. The presence of such spots can be inferred from
changes of the brightness, color and temperature. Thus, active stars
show variations in the brightness, color, and temperature. In
late-type stars chromospheric structures of stars are so closely
related to the magnetic field that chromospheric lines like
Ca\,II\,H,K are often used as a proxy for the magnetic field. For
example, Schrijver et al. (\cite{schrijver89}) derived a relation
between the flux in the emission core of these lines and the magnetic
flux. This relation has recently been studied in more detail by Rezaei et
al. (\cite{rezaei07}). Large-scale variations in the temperature and the
emission of chromospheric lines are common features of active stars.

We monitored two BDs over several rotation periods that are known to
have very strong magnetic fields of kG field strength. In contrast to
active stars, we find that the temperature variations are remarkably
small. In the case of \object{LP944-20} we find that the temperature
change over the surface is $\leq 30\,K$.  In other words, there cannot
be any large, cool spots on this object like on active stars.

Changes in the cloud-patterns would lead to changes in the equivalent
widths of spectral lines. Molina \& Moreno (\cite{molina92}), for
example, found changes of typically 30\% of the equivalent widths for
the strength of the $CH_4$ and $NH_3$ lines in different years and
different regions on Jupiter. In contrast to this, the rms-variations
of the K\,I-lines of \object{LP944-20} are only $\leq 2.0 \%$. Both
BDs must be remarkably homogeneous, with very little structure on
them.

 We recognize, however that $H_{\alpha}$ is now in emission, and there
 might even be some flare-like variability, demonstrating that this
 object is active. To study flares or inhomogeneities of the
 chromosphere, spectra of higher resolution are required.  The
 observation of flares in \object{LP944-20} would not be surprising,
 as huge flares have already been observed on this object in the X-ray
 regime (Rutledge et al. \cite{rutledge00}). The natural explanation
 for the absence of spots is that the coupling between the gas and the
 magnetic field is so low that the magnetic field does not create
 visible structures. The coupling is low because of the low
 temperature of the BDs. The same explanation cannot hold for the
 chromosphere because of its high temperature. As can be seen in
 Fig.\,\ref{fig04}, the variations of the pEW of $H_{\alpha}$ on time
 scales of a few hours are probably real but data of higher spectral
 resolution is needed to study the variations in detail. The relative
 size in the variations in the pEW would appear similar to what is
 observed on active stars.  However, we clearly did not observe a high
 level of flare-activity. We thus conclude that BDs, even with very
 strong magnetic fields, are unlike active stars, as they do not have
 prominent spots.  Since clouds are not affected by the magnetic
 fields, there is no reason why their structure should not change.
 Observationally, we find that the cloud structure remains
 remarkably constant.

Since it is shown in this paper that there are no significant
spectroscopic changes in the ultracool dwarfs \object{LP944-20} and
\object{2M0036+1821}, these objects can be considered as useful
spectroscopic standards for their spectral class. Our average spectra
will be made available to the community through the online IAC
ultracool dwarf catalog (Mart{\'{\i}}n, Cabrera \& Cenizo
\cite{martin05}).


\begin{acknowledgements}

We are grateful to the user support group of ESO/La Silla and the
2.2-m-telescope of the Centro Astron\'omico Hispano Alem\'an /CAHA) at
Calar Alto. This work made use of the SIMBAD database operated by the
CDS, France, and data products from the Two Micron All Sky Survey,
which is a joint project of the University of Massachusetts and the
Infrared Processing and Analysis Center/California Institute of
Technology, funded by the National Aeronautics and Space
Administration and the National Science Foundation. We also
acknowledge the use of the library of M, L, T dwarf spectra managed by
C.  Gelino, D. Kirkpatrick, A. Burgasser. We would also like to thank
the referee for helping us to improve the manuscript.

\end{acknowledgements}

\end{document}